# Anisotropic Optical Properties of 2D Silicon Telluride From Ab Initio Calculations

Romakanta Bhattarai and Xiao Shen[*]

Department of Physics and Materials Science, University of Memphis, Memphis, TN 38152, USA

**Abstract**

Silicon telluride ($Si_2Te_3$) is a silicon-based 2D chalcogenide with potential applications in optoelectronics. It has a unique crystal structure where Si atoms form Si-Si dimers to occupy the "metal" sites. In this paper, we report an ab initio computational study of its optical dielectric properties using the GW approximation and the Bethe-Salpeter equation (BSE). Strong optical anisotropy is discovered. The imaginary part of the dielectric constant in the direction parallel to the Si-Si dimers is found to be much lower than that perpendicular to the dimers. We show this effect originates from the particular compositions of the wavefunctions in the valence and conduction bands. BSE calculations reduce GW quasiparticle band gap by 0.3 eV in bulk and 0.6 eV in monolayer, indicating a large excitonic effect in $Si_2Te_3$. Furthermore, including electron-hole interaction in bulk calculations significantly reduces the imaginary part of the dielectric constant in the out-of-plane direction, suggesting strong interlayer exciton effect in $Si_2Te_3$ multilayers.

Since the last decade, two-dimensional materials have drawn a lot of interest in potential applications and fundamental sciences. Many 2D materials: graphene,[1-3] transition-metal dichalcogenides such as $MoS_2$,[4] phosphorene,[5-6] and more have been fabricated and the investigations revealed interesting properties different from their bulk forms. The unique opportunities in 2D systems include easy mechanical control,[7] rippling,[8] twisting,[9,10] easy chemical functionalization,[11,12] and defect engineering by irradiation.[13,14] Among the unique properties, the optical properties of 2D materials are particularity notable, as the low dimensionality significantly reduces dielectric screening, thus greatly enhancing the exciton binding energies.[15] As a result, 2D materials have attracted significant attention for optoelectronics.

Silicon telluride ($Si_2Te_3$) is a silicon-based two-dimensional material that has recently been made into a thickness of several atomic layers.[16] This material has a peculiar crystal structure: the Si atoms form Si-Si dimers to fill 2/3 of the allowed "metal" sites between the Te layers.[17, 18] There are four possible orientations of each Si-Si dimers, three in-plane, and one out-of-plane. The rotation of a Si-Si dimer has an activation energy of 1 eV and can happen at room temperature.[19] The Si-Si dimer orientation adds an additional internal degree of freedom that is unique in this material. The optical properties of $Si_2Te_3$ have been experimentally investigated for potential applications in optoelectronics. Photoluminescence measurements show that the band gap emission was observed below 90K, and the defect emission was observed at room temperature.[20] The relaxation of photocarriers also exhibits strong temperature dependence, possible related to the Si-Si dimer dynamics.[21] It was also demonstrated that the optoelectronic properties of the materials can be tuned by doping and intercalation.[22] For practical applications, $Si_2Te_3$ offers an additional advantage as it can be potentially compatible with the Si technology that is prevalent in the industry.[16]

---

[*] xshen1@memphis.edu

In this letter, we report a computational study of the optical properties of 2D $Si_2Te_3$. Using ab-initio many body GW approximation and Bethe-Salpeter equation (BSE), we obtain the dielectric constants of bulk and monolayer of $Si_2Te_3$. The results show strong optical anisotropy in $Si_2Te_3$. The imaginary part of the dielectric constant in the direction parallel to the Si-Si dimers is significantly smaller than the value perpendicular to the dimer. This effect is due to the compositions of the valence and conduction bands. We also find that including electron-hole interaction reduces quasiparticle band gap by 0.3 eV in bulk and 0.6 eV in a monolayer, indicating a large excitonic effect in $Si_2Te_3$. In addition, including electron-hole interaction in bulk calculations significantly reduces the imaginary part of the dielectric constant in the out-of-plane direction, indicating strong interlayer excitons in this material.

The calculations of the optical dielectric constants are carried out in three steps. In the first step, we perform standard density functional theory (DFT) calculations using the Perdew-Burke-Ernzerhof (PBE) exchange-correlation functional under generalized gradient approximation (GGA).[23] The pseudopotential used throughout the calculation was constructed under the projected augmented wave (PAW) method.[24] The electronic convergence is achieved when the energy difference between two successive steps is less than $10^{-9}$ eV. The atomic positions are fully relaxed until the energy difference between two successive steps is less than $10^{-8}$ eV. The integration over the Brillouin zone was performed with a grid of 3×3×3 k-point grid centered at Γ. Static and frequency-dependent dielectric constants are calculated by DFT method under the independent particle approximation. In the second step, we carry out many-body calculations using the GW approximation [25,26] that includes the quasiparticle correction to the DFT Kohn-Sham states.[27] In the GW approximation, the electronic self-energy (Σ) is approximated by a product of single-particle Green's function (G) and the screened Coulomb potential (W). In this work, single-shot GW ($G_0W_0$) calculations are performed. The calculation is done using 248 bands to take into account enough unoccupied bands. In the final step, we carry out calculations based on the Bethe-Salpeter equation (BSE), [28-30] which includes the electron-hole interaction (excitonic effect) that is absent in DFT and GW approaches. All the calculations were performed using the VASP (Vienna Ab initio Simulation Package) code.[31]

Figure 1 shows the relaxed structure of bulk $Si_2Te_3$. This configuration corresponds to the ground state of $Si_2Te_3$, where all the Si-Si dimers are oriented in the same direction.[19] We choose the y-axis to be along the direction of the Si-Si dimer. Although $Si_2Te_3$, like many other 2D chalcogenides, has a hexagonal-like lattice due to the packing of Te atoms, the primitive unit cell is actually triclinic because the two in-plane lattice vectors, a1 and a2, do not have the same length and their angle is not exactly 120°. Each bulk unit cell contains 2 vertially stacked monolayers. The primitive unit cell contains 8 Si atoms (4 Si-Si dimers) and 12 Te atoms.

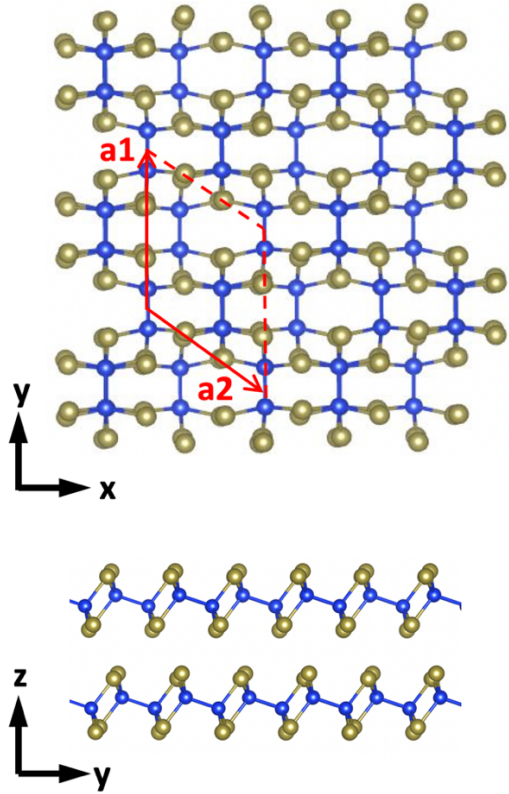

Figure 1. Top and side view of bulk $Si_2Te_3$. Si and Te atoms are in blue and tan color, respectively.

From DFT calculations, the static dielectric constant in the x-direction (perpendicular to dimers) is 9.92 while the static dielectric constant in the y-direction (parallel to the dimers) is 7.94. Considering the energy density in an electric field is $\frac{1}{2}\varepsilon E^2$, the 20% difference in static dielectric constants suggests that the energy in an electric field is lower when it is along the Si-Si dimer direction. As Si-Si dimer rotation has an activation energy of 1 eV and thus can happen at room temperature, under strong electric field, the dimers may align with the applied electric field to reduce the energy of the field. Therefore, a large electrical field may be used as a method to control the dimer alignment.

Figure 2 shows the imaginary part of the frequency-dependent dielectric constants of bulk $Si_2Te_3$ from DFT, GW, and BSE methods. From the DFT calculations, we deduce a band gap of 1.45 eV. From the GW calculations, we find the quasi-particle band gap to be 2.24 eV, much higher than the DFT result. From the BSE calculations, we obtain an excitonic band gap of 1.95 eV. The results suggest that the exciton binding energy is about 0.3 eV in bulk $Si_2Te_3$. Figure 3 shows the results of monolayer $Si_2Te_3$. The quasi-particle band gap from GW calculation is 2.85 eV, which is higher than the bulk value. From the BSE calculations, we obtain an excitonic band gap of 2.24 eV, suggesting a large exciton binding energy of 0.6 eV in monolayer Si$_2$Te$_3$.

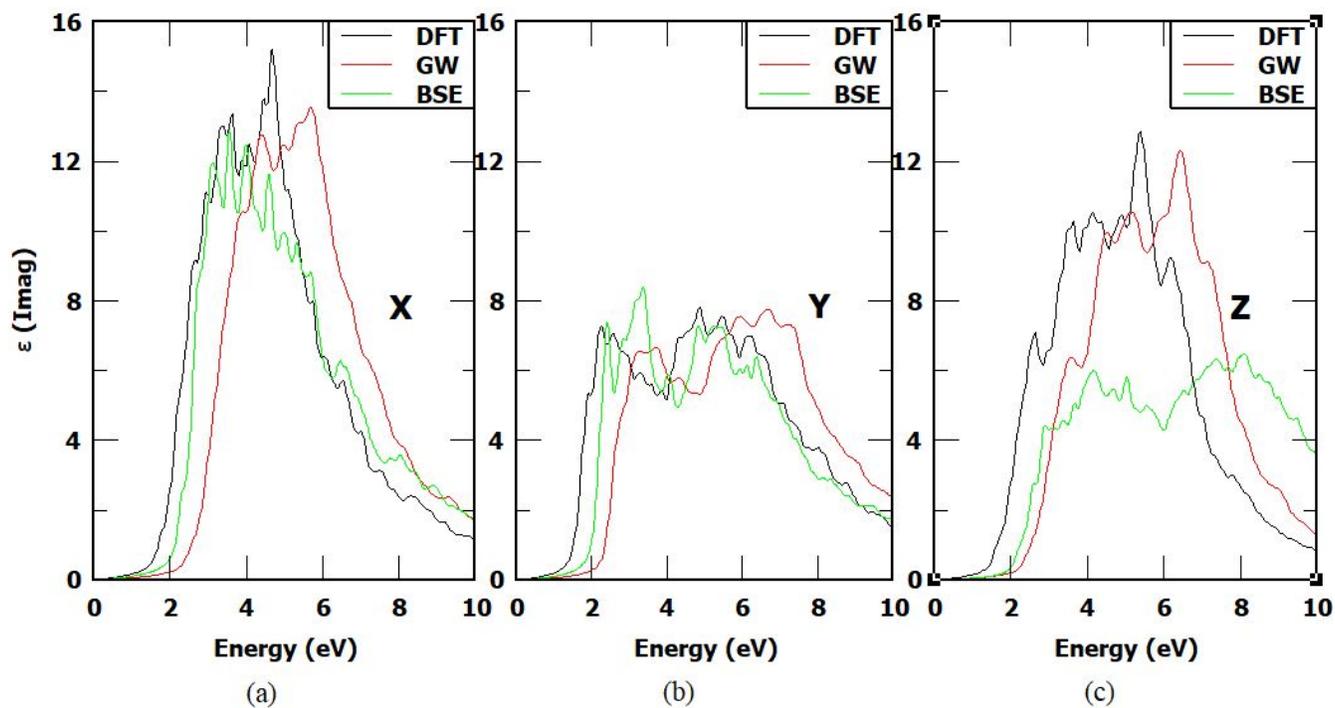

Figure 2. Imaginary part of the dielectric constant of bulk Si$_2$Te$_3$ along (a) x, (b) y, and (c) z-axes using three approaches (DFT, GW and BSE)

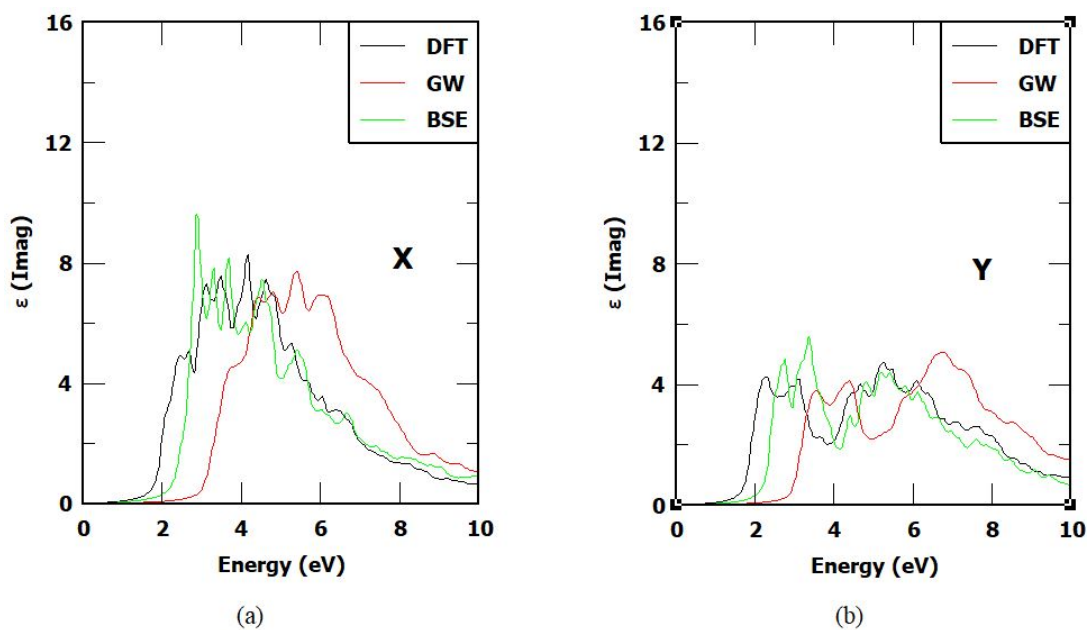

Figure 3. Imaginary part of the dielectric constant of monolayer Si$_2$Te$_3$ along (a) x, (b) y, axes using three approaches (DFT, GW and BSE).

Strong anisotropy in the frequency-dependent dielectric constants is observed in both bulk and monolayer. From BSE calculations of bulk $Si_2Te_3$, the maximum $Im(\varepsilon_x)$ is observed at 3.3 eV of photon energy, with a peak value of 13.0 (Figure 2a). Meanwhile, the maximum $Im(\varepsilon_y)$ is observed at a similar photon energy, with a peak value of 8.5 (Figure 2b). The results indicate that $Si_2Te_3$ is highly anisotropy in the optical regime. The fact that $Im(\varepsilon_x)$ is significantly larger than $Im(\varepsilon_y)$ in optical regime is also observed in DFT and GW calculations, and in the case of 2D monolayers as well (Figure 3). This anisotropy originates from the specific composition of the conduction and valence bands. In Figure 4, we show the wavefunctions of the conduction band minimum (CBM) and valence band maximum (VBM) in bulk $Si_2Te_3$. The oscillation strength under electric field in x or y-direction is proportional to $\langle \phi_{CB}|x|\phi_{VB}\rangle$. As can be seen in Figure 4a, the VBM contains the Te 5p orbitals, one of which is marked as point 1. Meanwhile, for CBM (Figure 3c), the Si 2s orbitals are part of the wavefunction as marked as point 2 and 3. Under an electric dipole perturbation in the x-direction, this orbital at point 1 in VBM will expand horizontally and have a larger overlap with wavefunction at points 2 and 3 CBM, resulting in a large oscillation strength. For electric dipole perturbation in the y-direction, the enhancement of the oscillation strength is not significant.

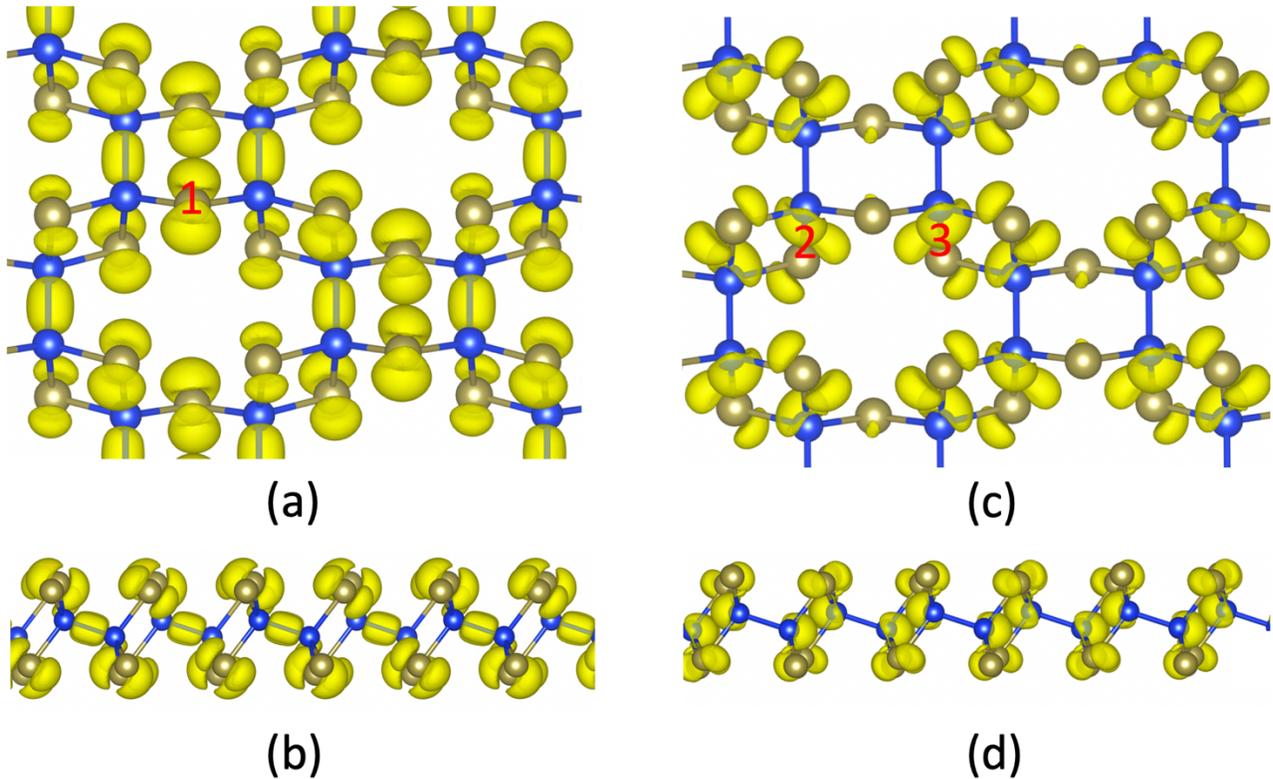

Figure 4: Top view (a) and side view (b) of the module-squared wavefunctions at VBM of bulk $Si_2Te_3$. Top view (c) and side view (d) of the module-squared wavefunctions at CBM. Each bulk unit cell contains two layers of $Si_2Te_3$, only one layer is shown here for clarity.

Figure 2c shows an additional intriguing feature: the imaginary part of the dielectric constant in the vertical direction, Im($\varepsilon_z$), is significantly smaller in BSE calculations than in DFT and GW calculations. This result indicates that the electron-hole interaction has a strong effect on the spectra of the quasi-electron and the quasi-hole when their wavevectors differ in the z-direction, which suggests a strong interlayer excitonic effect [32,33] in the optical properties of $Si_2Te_3$ multilayers. The strong interlayer excitonic effect can be understood from the particular compositions of the wavefunctions at the CBM and VBM as well. As shown in Figure 4, the VBM consists of Te 5p orbitals and the bonding orbitals of Si atoms, while the CBM of $Si_2Te_3$ consists of the Te 5p orbitals and the anti-bonding orbitals on Si atoms. From the side views of the wavefunctions (Figure 4b and 4d), it is obvious that both CBM and VBM contain Te 5p orbitals that extend into the space between the $Si_2Te_3$ layers. Therefore, the wavefunction of an electron in one layer and that of a hole in an adjacent layer can be closely spaced. The close proximity leads to large Coulombic interactions and naturally strong interlayer exciton effect.

In Summary, we report a computational study of the optical dielectric properties of $Si_2Te_3$ using the GW and BSE approaches. The material exhibits strong optical anisotropy. The imaginary part of the dielectric constant in the direction parallel to the Si-Si dimers is much smaller compared to the direction perpendicular to the dimer. The electron-hole interaction reduces the quasiparticle band gap by 0.3 eV in bulk and 0.6 eV in the case of a monolayer, indicating a large excitonic effect in $Si_2Te_3$. Besides, BSE calculation significantly reduces the imaginary part of the dielectric constant of bulk $Si_2Te_3$ in the out-of-plane direction, suggesting a strong Coulombic interaction in the case of interlayer excitons.

**ACKNOWLEDGEMENTS**

This work was supported by National Science Foundation grant # DMR 1709528 and by the Ralph E. Powe Jr. Faculty Enhancement Awards from Oak Ridge Associated Universities (ORAU). Computational recourses were provided by University of Memphis High-Performance Computing Center (HPCC) and by the NSF XSEDE under grants # TG-DMR 170064 and 170076. We thank Dr. Li Yang, Dr. Jingbiao Cui, and Dr. Thang B. Hoang for helpful discussion.